\documentclass[showpacs,preprintnumbers,amsmath,amssymb]{revtex4}
\usepackage{graphicx}
\usepackage{dcolumn}
\usepackage{bm}
\usepackage{graphicx}
\usepackage{amssymb}
\usepackage{amsmath}
\begin{document}

\title{Low-lying spectrum of the Y-string three-quark potential using hyper-spherical coordinates}
\author{V. Dmitra\v sinovi\' c,\\
{\it 
Vin\v ca Institute of Nuclear Sciences, Physics laboratory 010} \\
{\it P.O.Box 522, 11001 Beograd, Serbia} \\
Toru Sato,\\
{\it Dept. of Physics, Graduate School of Science, Osaka University}\\
{\it Toyonaka 560-0043, Japan} \\
Milovan {\v S}uvakov \\
{\it Institut J. Stefan, Dept. of Theoretical Physics F-1,}\\
{\it  Jamova 39, 1000 Ljubljana, Slovenia}}
\date{\today}
%
\begin{abstract}
We calculate the energies of three-quark states with definite
permutation symmetry (i.e. of SU(6) multiplets) in the N=0,1,2
shells, confined by the Y-string three-quark potential. The exact
Y-string potential consists of one, so-called three-string term,
and three angle-dependent two-string terms. Due to this technical
complication we treat the problem at three increasingly accurate
levels of approximation: 1) the (approximate) three-string
potential expanded to first order in trigonometric functions of
hyper-spherical angles; 2) the (approximate) three-string
potential to all orders in the power expansion in hyper-spherical
harmonics, but without taking into account the transition(s) to
two-string potentials; 3) the exact minimal-length string
potential to all orders in power expansion in hyper-spherical
harmonics, and taking into account the transition(s) to two-string
potentials. We show the general trend of improvement 
of these approximations: The exact non-perturbative corrections to
the total energy are of the order of one per cent, as compared
with approximation 2), yet the exact energy differences between
the $[20,1^{+}], [70,2^{+}], [56,2^{+}], [70,0^{+}]$-plets are
shifted to 2:2:0.9, from the Bowler and Tynemouth separation rule
2:2:1, which is obeyed by approximation 2) at the one per cent
level. The precise value of the energy separation of the first
radial excitation (``Roper") $[56^{\prime},0^{+}]$-plet from the
$[70,1^{-}]$-plet depends on the approximation, but does not
become negative, i.e. the ``Roper" remains heavier than the
odd-parity $[70,1^{-}]$-plet in all of our approximations.
\end{abstract}


\maketitle


\section{Introduction}

\begin{figure}[tbp]
\centerline{\includegraphics[width=2.5in,,keepaspectratio]{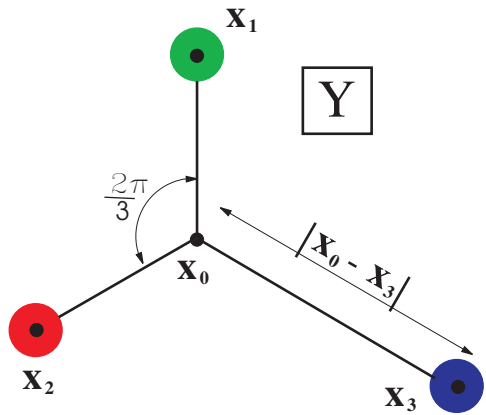}}
\caption{Three-quark Y-junction string potential.} \label{Y_3q}
\end{figure}

\begin{figure}[tbp]
\centerline{\includegraphics[width=2.5in,,keepaspectratio]{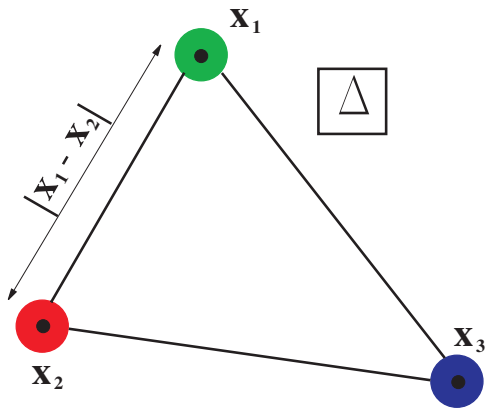}}
\caption{Three-quark $\Delta$-shape string potential.}
\label{D_3q}
\end{figure}

The so-called Y-junction string three-quark potential, defined by
\begin{equation}
\label{conf_Y} V_Y = \sigma \min_{\bf x_0}\; \sum_{i=1}^3 |{\bf
x_i} - {\bf x_0}|.
\end{equation}
has long been advertised
\cite{artr75,dosc76,carl83,caps86,blas90,koma01} as the natural
approximation to the flux tube confinement mechanism, that is
allegedly active in QCD. Lattice investigations, Refs.
\cite{bali00,taka01,Alex01}, however, contradict each other in
their attempts to distinguish between the Y-string, Fig.
\ref{Y_3q}, and the $\Delta$-string potential, see Fig.
\ref{D_3q},
\begin{equation}
\label{conf_D} V_{\Delta} = \sigma \sum_{i<j =1}^3 |{\bf x}_{i} -
{\bf x}_{j}|,
\end{equation}
which, in turn, is indistinguishable from the sum of three linear
two-body potentials. The present point of view held among the
lattice QCD practitioners is that there should be a smooth
cross-over from the $\Delta$ to the Y-potential at interquark
distances of around 0.8 fm \cite{Caselle:2005sf}. Exactly how this
cross-over should be implemented is not clear just now. Moreover,
quantum fluctuations of the three flux tubes lead to L{\" u}scher
type corrections in the potential \cite{deForcrand:2005vv}. The
evaluation of such corrections is beyond the scope of the present
paper.

Over the past 25 years, the Y-string potential has been used in
several, more or less systematic studies of baryons in the
(constituent) quark model with various hyperfine interactions
\cite{carl83,caps86,blas90,Bonn01}, and yet some of the most basic
predictions of this potential acting alone, i.e. without hyperfine
interaction, such as the spectrum of the low-lying three-quark
states remain unknown. The older calculations \cite{carl83,caps86}
treat the Y-string potential only in first-order perturbation
theory, whereas the more recent ones \cite{Bonn01} invoke
equivalence with $\Delta$ string potential\footnote{``the Y-string
and the $\Delta$-string potentials are essentially
indistinguishable"}, up to an overall multiplicative factor $f
\simeq 0.5493$ ($\frac12 < f < \frac{1}{\sqrt3}$) for the string
tension $\sigma$. That leads us to a third set of calculations
that rely on the (as yet not proven) ``equivalence" of the
Y-string potential and the $\Delta$ string one to look for the
``best two-body approximants" to the Y-string potential, see Refs.
\cite{fabr97,Grenoble03} and follow-up references. Note, however,
that these papers use only the hyper-radial approximation and they
are only concerned with the overall strength of the coupling.

We wish to resolve this dilemma using analytical methods, whenever
possible \footnote{It would be ideal if these two kinds of string
potentials were to predict (dramatically) different baryon spectra
that can be easily differentiated by experiment, but that turns
out not to be the case}. So, at least for the time being, we have
to resort to calculations in the nonrelativistic quark model. We
shall not repeat the two-body $\Delta$ string potential
calculation, as there is a large body of literature on the
subject: instead, we shall use the paper by J.M. Richard and P.
Taxil \cite{RT90} as a benchmark. That paper uses the
hyper-spherical harmonics method, which we also use, with only a
slightly different practical implementation: whereas Ref.
\cite{RT90} uses the Jacobi coordinates (see Sect.
\ref{s:potential}) in the evaluation of the relevant matrix
elements, we shall evaluate them directly in the hyperspherical
variables.

For the sake of completeness and clarity we shall start with two
extensive technical preparatory sections: First, in Sect.
\ref{s:potential} we outline the three main technical problems,
and then in Sect. \ref{s:method} we address them one after
another; for this reason we divide Sect. \ref{s:method}, in which
we formulate the methods that we use, into three parts: the first
one, Sect. \ref{s:ang}, is about the angular momentum recoupling
algebra necessary to deal with the non-conserved ``partial"
angular momenta, the second part, Sect. \ref{s:pert}, deals with
the square root(s) in the Y-string potential, and finally the
third one, Sect. \ref{s:nonpert}, addresses all four forms of the
string potential together. Our results are shown in Sect.
\ref{s:results}, which is divided into four parts: one part that
comprises Sect. \ref{s:pert-result} deals with the first-order
perturbative approximation results, wherein the angular dependence
is treated as a (small) perturbation to the linear potential
oscillator, Sect. \ref{s:HSY-D} and two with non-perturbative ones
in Sect. \ref{s:allorder-result}: one for the Y-string potential
in Sect. \ref{s:HSY-D1}, and another for the complete string
potential in Sect. \ref{s:2string-result}. The final Section
\ref{s:summary} contains a summary and discussion of our results.
A number of technical issues are discussed in Appendices
\ref{s:Y-HSpert},\ref{s:Y-HS ME},\ref{s:Hrad}.

\section{The Y-string potential}
\label{s:potential}

The complexity of the potential Eq. (\ref{conf_Y}) is best seen
when expressed in terms of three-body Jacobi (relative)
coordinates ${\bm \rho},{\bm \lambda}$
\begin{eqnarray}
{\bm \rho} &=& \frac{1}{\sqrt{2}}({\bf x_1} - {\bf x_2}),
\label{e:rho} \\ 
{\bm \lambda}&=& \frac{1}{\sqrt{6}}({\bf x_1} + {\bf x_2}- 2 {\bf
x_3}), \label{e:lambda} \
\end{eqnarray}
as follows. The exact string potential  Eq. (\ref{conf_Y})
consists of the so-called Y-string, or three-string term,
\begin{eqnarray}
V_{\rm string} &=& V_{\rm Y} = \sigma \sqrt{\frac{3}{2}({\bm
\rho}^2 + {\bm \lambda}^2 + 2 |{\bm \rho} \times {\bm
\lambda}|)} ,\label{hypVY1a} \\
&& \textrm{when}  \left\{
\begin{array}{l}
\quad 2 {\bm \rho}^2 - \sqrt{3}{\bm \rho} \cdot {\bm \lambda} \geq
- \rho \sqrt{{\bm \rho}^2 + 3 {\bm \lambda}^2 - 2 \sqrt{3}{\bm
\rho} \cdot {\bm \lambda}} \nonumber \\
\quad 2 {\bm \rho}^2 + \sqrt{3}{\bm \rho} \cdot {\bm \lambda} \geq
- \rho \sqrt{{\bm \rho}^2 + 3 {\bm \lambda}^2 + 2 \sqrt{3}{\bm
\rho} \cdot {\bm \lambda}}  \\
\quad 3 {\bm \lambda}^2 - {\bm \rho}^2 \geq - \frac{1}{2} \sqrt{
({\bm \rho}^2 + 3 {\bm \lambda}^2)^{2} - 12 ({\bm \rho} \cdot {\bm
\lambda})^{2}}
\, .\\
\end{array} \right. \
\end{eqnarray}
and three angle-dependent two-part string, or the so-called
V-string, terms,
\begin{subequations}
\begin{eqnarray}
V_{\rm string} &=& \sigma  \left( \sqrt{\frac{1}{2}({\bm \rho}^2 +
3 {\bm \lambda}^2 + 2 \sqrt{3}{\bm \rho} \cdot {\bm \lambda})} +
\sqrt{\frac{1}{2}({\bm \rho}^2 + 3 {\bm \lambda}^2 - 2
\sqrt{3}{\bm \rho} \cdot {\bm \lambda})}\right) \label{hypVY1b} \\
&& \textrm{when}  \left\{
\begin{array}{l}
\quad 2 {\bm \rho}^2 - \sqrt{3}{\bm \rho} \cdot {\bm \lambda} \geq
- \rho \sqrt{{\bm \rho}^2 + 3 {\bm \lambda}^2 - 2 \sqrt{3}{\bm
\rho} \cdot {\bm \lambda}} \nonumber \\
\quad 2 {\bm \rho}^2 + \sqrt{3}{\bm \rho} \cdot {\bm \lambda} \geq
- \rho \sqrt{{\bm \rho}^2 + 3 {\bm \lambda}^2 + 2 \sqrt{3}{\bm
\rho} \cdot {\bm \lambda}}  \\
\quad 3 {\bm \lambda}^2 - {\bm \rho}^2 \leq - \frac{1}{2} \sqrt{
({\bm \rho}^2 + 3 {\bm \lambda}^2)^{2} -
12 ({\bm \rho} \cdot {\bm \lambda})^{2}}\\
\end{array} \right. \\
V_{\rm string} &=& \sigma \left( \sqrt{2} \rho +
\sqrt{\frac{1}{2}({\bm \rho}^2 + 3 {\bm \lambda}^2 + 2
\sqrt{3}{\bm \rho} \cdot {\bm \lambda})} \right)
\label{hypVY1c} \\
&& \textrm{when}  \left\{
\begin{array}{l}
\quad 2 {\bm \rho}^2 - \sqrt{3}{\bm \rho} \cdot {\bm \lambda} \geq
- \rho \sqrt{{\bm \rho}^2 + 3 {\bm \lambda}^2 - 2 \sqrt{3}{\bm
\rho} \cdot {\bm \lambda}} \nonumber \\
\quad 2 {\bm \rho}^2 + \sqrt{3}{\bm \rho} \cdot {\bm \lambda} \leq
- \rho \sqrt{{\bm \rho}^2 + 3 {\bm \lambda}^2 + 2 \sqrt{3}{\bm
\rho} \cdot {\bm \lambda}} \nonumber \\
\quad 3 {\bm \lambda}^2 - {\bm \rho}^2 \geq - \frac{1}{2} \sqrt{
({\bm \rho}^2 + 3 {\bm \lambda}^2)^{2} - 12 ({\bm \rho} \cdot {\bm
\lambda})^{2}}
\end{array} \right. \\
V_{\rm string} &=& \sigma \left( \sqrt{2} \rho +
\sqrt{\frac{1}{2}({\bm \rho}^2 + 3 {\bm \lambda}^2 - 2
\sqrt{3}{\bm \rho} \cdot {\bm \lambda})}\right)
\label{hypVY1d} \\
&& \textrm{when}  \left\{
\begin{array}{l}
\quad 2 {\bm \rho}^2 - \sqrt{3}{\bm \rho} \cdot {\bm \lambda} \leq
- \rho \sqrt{{\bm \rho}^2 + 3 {\bm \lambda}^2
- 2 \sqrt{3}{\bm \rho} \cdot {\bm \lambda}} \\
\quad 2 {\bm \rho}^2 + \sqrt{3}{\bm \rho} \cdot {\bm \lambda} \geq
- \rho \sqrt{{\bm \rho}^2 + 3 {\bm \lambda}^2 + 2 \sqrt{3}{\bm
\rho} \cdot {\bm \lambda}}  \nonumber \\
\quad 3 {\bm \lambda}^2 - {\bm \rho}^2 \geq - \frac{1}{2} \sqrt{
({\bm \rho}^2 + 3 {\bm \lambda}^2)^{2} - 12 ({\bm \rho} \cdot {\bm
\lambda})^{2}}
\end{array} \right.
\, .
\end{eqnarray}
\end{subequations}
Here, the reasons for the lack of use of the exact potential Eq.
(\ref{conf_Y}) become clear: i) it is a genuine three-body
operator with a complicated and unusual (``area term") angular
dependence under the square-root of the most important term (the
Y-junction string potential) that leads to the non-conservation of
the individual Jacobi coordinates' angular momenta \cite{SD07} and
hugely complicates the equations of motion; ii) the square-roots
appearing in all four functional forms of the potential make this
task even more difficult; iii) the presence of four different
functional forms of the potential depending on the configuration
space angles makes the integration of the equations of motion
difficult as one cannot easily separate the angular and radial
integrals.

Perhaps the simplest, yet realistic approximation to the exact
string potential Eq. (\ref{conf_Y}) is the Y-string, or the
``three-string" potential, Eq. (\ref{hypVY1a}), that is used in
the whole configuration space, i.e., even when one of the angles
in the triangle exceeds $120^{\circ}$. In that way one avoids the
cumbersome transition to the V-string potentials, see problem iii)
above \footnote{It has been claimed that this approximation is
exact \cite{fabr97}, however, in the case(s) of zero individual
orbital angular momenta $\left(l_{\rho}, l_{\lambda}\right)$, i.e.
for the radial excitations of the S-wave ground state, and
presumably a good one for low values of the orbital angular
momenta. We shall show below that this claim is incorrect,
however.}. Still, even this simplified approximation suffers from
two difficulties mentioned above: i) an unusual (``area term")
angular dependence under the square-root that leads to the
non-conservation of the individual Jacobi coordinates' angular
momenta; ii) the square-root. We shall address these problems in
successive steps: i) the area term turns out to be exactly
(analytically) integrable, but it requires complicated angular
momenta recoupling algebra and the exact value of a particular one
dimensional angular integral. Problem ii), the square root, can be
treated, at first, by a series expansion, i.e. in perturbation
theory, and then by numerical evaluation of the complete
functional expression, i.e. in non-perturbative approximation.
Finally, the last issue iii) is tackled, at the price of
considerable inconvenience: the necessary hyperangular matrix
elements of the complete three-body potential can be evaluated
using above mentioned methods, except that two previously separate
integrals (over one quasi-radial and one angular variable) have to
be performed simultaneously, as the integration boundaries involve
both variables.

It turns out that the crucial ingredient for the success of this
effort is the application of the so-called hyper-spherical
coordinates/angles \cite{Simonov:1965ei}, or a particular
variation thereof, the cosines of the relative angle $\theta$
between the Jacobi coordinates ${\bm \rho},{\bm \lambda}$ and of
the angle $2 \chi$ defined by way of the ratio of the moduli
${\rho}, {\lambda}$ of the two Jacobi coordinates.

The hyper-spherical method has been used widely in the few-body
atomic and nuclear physics, for review see Ref. \cite{Krivec}, but
we are aware of only one paper, Ref. \cite{fabr97}, that uses it
in the context of the three-quark Y-string problem, which is
perhaps ironic, as it turns out to be the most natural set of
coordinates for the problem at hand, whereas its use in two-body
atomic and nuclear potential problems is plagued by slow
convergence of the expansion. Some papers have applied the
hyper-radial approximation to the three-quark problem
\cite{Grenoble03,RT90}, but that misses the essential points
discussed here.

\section{Methods}
\label{s:method}

The first problem (angular momentum) is generic to all (string)
three-body potentials and is thus independent of the approximation
used, so its solution will be used subsequently in both
perturbative and non-perturbative approximations. Moreover, its
solution can be incorporated into the hyper-spherical formalism
that will used later. For this reason we start with the angular
momentum recoupling.

\subsection{Angular momentum recoupling}
\label{s:ang}

The `vector cross-product, or the ``area term" $2\, |{\bm \rho}
\times {\bm \lambda}\, |$ in this potential has some curious
properties in classical and quantum mechanics: it conserves the
sum ${\bf L} = {\bf l}_{\rho} + {\bf l}_{\rho}$ of the two partial
(orbital) angular momenta ${\bf l}_{\rho} = {\bm \rho} \times {\bf
p_{\rho}}$ and ${\bf l}_{\lambda}= {\bm \lambda} \times {\bf
p_{\lambda}}$, but not their difference \cite{SD07}, i.e., the
individual (orbital) angular momenta are not conserved. As a
consequence, there can be ``spilling" of the orbital angular
momentum from the $\rho$ (normal) mode into the $\lambda$ one and
{\it vice versa}. Only the radial excitations of the S-wave ground
state remain immune to this spillage.

We need to evaluate matrix elements of the following form
\begin{eqnarray}
\langle \left(l_{1f} \otimes l_{2f}\right) L_{f},M_{f}| |\sin
\theta | |\left(l_{1i} \otimes l_{2i}\right) L_{i},M_{i} \rangle
&=& \sum_{m_{1f},m_{2f},m_{1i},m_{2i}}(l_{1f}, m_{1f} ; l_{2f},
m_{2f} | L_f, M_f)
\nonumber \\
(l_{1i} ,m_{1i}; l_{2i} ,m_{2i}|L_i,M_i)
& \times& \langle l_{1f} ,m_{1f}; l_{2f} ,m_{2f} | |\sin \theta |
|l_{1i} ,m_{1i}; l_{2i} ,m_{2i} \rangle \ \label{e:sinME}
\end{eqnarray}
where
\begin{eqnarray}
|\, {\hat \rho} {\times} {\hat \lambda} \,|\,  = \frac{|\, {\bm
\rho} \times {\bm \lambda} \,|}{\rho \lambda} = | \sin \theta | =
\sqrt{1 - \cos^{2}\theta}
\end{eqnarray}
and
\begin{eqnarray}
\cos\theta &=& \cos\theta_{1} \, \cos\theta_{2} + \cos(\varphi_{1}
- \varphi_{2})\, \sin\theta_{1}\, \sin\theta_{2}
\nonumber \\
&=& -\left( \,\frac{4\,\pi}{\sqrt{3}}\right)
\left[Y_{1}(\theta_1,\phi_1) \otimes Y_{1}(\theta_2,\phi_2)
\right]_{L=0}
\end{eqnarray}
that is to be inserted into
\begin{eqnarray}
\langle l_{1f},m_{1f}, l_{2f},m_{2f} | |\sin \theta |
|(l_{1i},m_{1i}, l_{2i},m_{2i})\rangle &=&
\nonumber \\
\int d\Omega_1 \int d\Omega_2 Y^*_{l_{1f},m_{1f}}(\theta_1,\phi_1)
Y_{l_{1i},m_{1i}}(\theta_1,\phi_1)
Y^*_{l_{2f},m_{2f}}(\theta_2,\phi_2)
Y_{l_{2i},m_{2i}}(\theta_2,\phi_2) &\times& |\sin \theta | \
\end{eqnarray}
Capstick and Isgur, Ref. \cite{caps86} have reduced the angular
matrix elements Eq. (\ref{e:sinME}) to
\begin{eqnarray}
\langle (l_{1f}\otimes l_{2f})L_f,M_f||\sin\theta||(l_{1i}\otimes
l_{2i})L_i,M_i \rangle &=& \delta_{L_f,L_i}\delta_{M_f,M_i}
\sum_{L} \sqrt{(2l_{1f}+1)(2l_{2f}+1)(2l_{1i}+1)(2l_{2i}+1)}
\nonumber \\
&\times& (-1)^{L+l_{1f}+l_{2f}}
W(l_{1i},L,L_i,l_{2f};l_{1f},l_{2i})
(l_{1f},0,l_{1i},0|L,0)(l_{2f},0,l_{2i},0|L,0)
 \nonumber \\
& \times & \frac{1}{2}\int_{-1}^{1}d\cos\theta P_L(\cos \theta)
|\sin\theta|
\end{eqnarray}
a sum of products of SU(2) Clebsch-Gordan
$(l_{1f},0,l_{1i},0|L,0)(l_{2f},0,l_{2i},0|L,0)$ and Racah
$W(l_{1i},L,L_i,l_{2f};l_{1f},l_{2i})$ coefficients, which can be
found in standard collections of angular momentum tables, such as
Ref. \cite{VK89}, and an integral over even-$L$ order Legendre
polynomials $P_{L}(x)$ that can be reduced to a ratio of
double-factorial functions \cite{Iwanami}:
\begin{eqnarray}
\frac{1}{2}\int_{-1}^{1}d\cos\theta P_L(\cos \theta) \sqrt{1 -
\cos^2\theta}  &=&~~~ \left \{
\begin{array}{l} ~~~0
  \ ,\ \  \mbox{for} \ \ \mbox{odd}~L = 1,3,5,... \\
~~~\frac{\pi}{4} , \ \  \mbox{for} \ \ \ L = 0 \\
- \frac{\pi}{2L}\frac{(L-1)!!(L-3)!!}{(L+2)!!(L-2)!!}
  \ , \ \ \mbox{for} \ \  \mbox{even}~L = 2,4,6, ... \ne 0 \\
\end{array}\right.
\label{e:int}
\end{eqnarray}
In Table \ref{tab:drcc1a} we show the dependence of these matrix
elements on the ``partial" orbital angular momenta $l_{\rho},
l_{\lambda}$ - note in particular the non-vanishing off-diagonal
matrix element that ensures the existence of the mass-splitting
``mixing" terms in the $[56,2^+]$ and the $[70,2^+]$-plets. Note
that all of the angular matrix elements are exact integer or
fractional multiples of the basic unit $(\frac{\pi}{4})$, thus
leading to the analytic form of the angular matrix elements, shown
in the eighth column of Table \ref{tab:drcc1a}.
\begin{table}[tbh]
\begin{center}
\caption{Non-vanishing diagonal and off-diagonal matrix elements
$\langle |\sin \theta| \rangle = \langle \left(l_{1f} \otimes
l_{2f}\right)_{L_{f}}^{M_{f}}| \,|\sin \theta | \, |\left(l_{1i}
\otimes l_{2i}\right)_{L_{i}}^{M_{i}} \rangle$, in total S-, P-
and D-wave states, and partial waves $l_{1f} = l_{2f} = l_{1i} =
l_{2i} = 0,1,2$.}
\begin{tabular}{cccccc}
\hline \hline N & $L_f$ & $l_{\rho f} \times l_{\lambda f}$ &
$L_i$ & $l_{\rho i} \times l_{\lambda i}$ & $\frac{4}{\pi} \langle
|\sin \theta| \rangle$
\\
 \hline
0 & 0 & $0\times0$ & 0 & $0\times0$ & 1 \\
2 & 0 & $1\times1$ & 0 & $1\times1$ & $\frac{3}{4}$ \\
1 & 1 & $1\times0$ & 1 & $1\times0$ & 1 \\
2 & 1 & $1\times1$ & 1 & $1\times1$ & $\frac{9}{8}$ \\
2 & 2 & $1\times1$ & 2 & $1\times1$ & $\frac{39}{40}$\\
2 & 2 & $2\times0$ & 2 & $2\times0$ & 1\\
2 & 2 & $2\times0$ & 2 & $0\times2$ & -$\frac{1}{8}$ \\
\hline \hline
\end{tabular}
\label{tab:drcc1a}
\end{center}
\end{table}

\subsection{Perturbative approximation}
\label{s:pert}

We must first determine the qualitative features of the ``area
term" $2\, |{\bm \rho} \times {\bm \lambda}\, |$ and its angular
momentum dependence on the ordering of states in quantum
mechanics, e.g. whether the S-wave states, such as the Roper, are
lowered or raised in energy as compared with other states in the
N=2 band, e.g. the P- and D-waves?

A historical remark seems in order now: It has been known at least
since the late 1970's, see Refs. \cite{IK79,GS79}, that arbitrary
an-harmonic (both two- and three-body) potentials split the N=2
shell harmonic oscillator states\footnote{but not the N=0,1 shells
states.} according to their spatial permutation symmetry classes,
or, what is the same, according to the SU(6) multiplets. This
splitting of the harmonic oscillator spectrum has been worked out
to various degrees of mathematical sophistication in Refs.
\cite{GS79,IK79,BT83,RT90} for arbitrary two-body potentials, but
has been only briefly mentioned in the case of three-body ones
\cite{BT83}. In particular Bowler and Tynemouth \cite{BT83} have
shown, on the basis of the Sp(12,R) group theory and first-order
perturbation theory, that the energy ordering and splitting of
four of the five SU(6) multiplets in the N=2 band always remain
the same for arbitrary permutation-symmetric three-body
potentials, the only exception being the energy of the $[56,0^+]$
multiplet (containing the Roper resonance), which remains
unconstrained by this theorem. In other words, the key to the
mystery of the Roper's abnormally low mass may well reside in the
form of the three-quark potential.

\subsubsection{The square root in the Y-string potential}
\label{s:root}

\begin{figure}[tbp]
\centerline{\includegraphics[width=2.5in,,keepaspectratio]{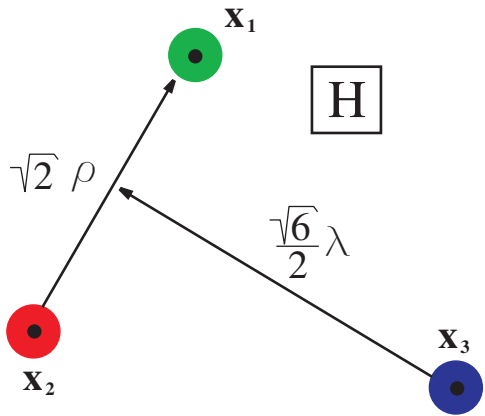}}
\caption{Two three-body Jacobi coordinates $\rho, \lambda$ that
define the ``hyper-radial" string length $\sqrt{|{\bf \rho}|^2 +
|{\bf \lambda}|^2}$.} \label{H_3q}
\end{figure}

Perhaps the simplest way to address these questions is to expand
the root in the Y-string potential Eq. (\ref{hypVY1a}) in a power
series
\begin{eqnarray}
V_{\rm Y} &=& \sigma \sqrt{\frac{3}{2}({\bm \rho}^2 + {\bm
\lambda}^2)} \left(1 + \frac{|{\bm \rho} \times {\bm
\lambda}|}{{\bm \rho}^2 + {\bm \lambda}^2} + \cdots \right).\
\label{e:expans}
\end{eqnarray}
and then, keep only a few lowest-order non-trivial terms, apply
the first-order perturbation theory. The unperturbed potential in
Eq. (\ref{e:expans}) is slightly more complicated than the
harmonic oscillator, so we break this into two steps:

We shall use the quasi-linear hyper-radial potential to define the
unperturbed Hamiltonian, which leads us to introduce the
``hyper-spherical (H) string length" $l_{\rm H}$, or the
``hyper-radius" $R$, as follows
\begin{equation}
R = \sqrt{\rho^{2} + \lambda^{2}} = l_{\rm H}.
\end{equation}
The two Jacobi vectors ${\bf \rho}, {\bf \lambda}$ defined in Eqs.
(\ref{e:rho}),(\ref{e:lambda}), are shown in Fig. \ref{H_3q}.
Instead of two three-vectors $\rho$ and $\lambda$, the
hyper-spherical formalism introduces the hyper-radius $R$ and the
(new) hyper-angle $\chi$ by way of the ``polar transformation"
\begin{equation}
\rho = R \sin \chi,\quad \lambda = R \cos \chi \quad \text{with}
\quad 0 \le \chi \le \pi/2.
\end{equation}
Thus we see that the expansion in Eq. (\ref{e:expans}) corresponds
precisely to the expansion in powers of trigonometric functions of
hyper-angles:
\begin{eqnarray}
V_{\rm Y} &=& \sigma \sqrt{\frac{3}{2} R^2 (1 + \sin 2 \chi |\sin
\theta|)} ,
\nonumber \\
&=& \sigma \sqrt{\frac{3}{2}} R \left(1 + \frac12 \sin 2 \chi
|\sin \theta| + \cdots \right) , \label{e:exp1} \
\end{eqnarray}
the so-called hyper-spherical expansion, the first term being the
``hyper-radial" term and the rest corresponding to higher-order
``hyper-spherical harmonics".

The excited three-quark states then naturally fall into
(non-degenerate) multiplets (labelled by the ``grand angular
momentum" quantum number K) of the broken O(6) symmetry, whose
degeneracy is broken by the hyper-angular dependent perturbation,
see Appendix \ref{s:Y-HSpert}. These members/subsets of the
(broken) O(6) multiplets coincide with the ``ordinary" SU(6)
multiplets of standard quark model in the two lowest lying bands
(usually denoted by the harmonic oscillator label N=0,1), the only
distinction being that in the N=2 band the first radial excitation
of the N=0 ground state, a.k.a. the ``Roper", does not fall into
the K=2 band, but rather belongs to the K=0, hyper-radial
excitation ${\cal N}_{\rm K} = 2$ ``shell" consisting of exactly
one state, see Appendix \ref{s:K012}. Thus, the symmetrized
hyper-spherical harmonics are the most natural tool to describe
the non-relativistic three-quark SU$_{FS}$(6) multiplets' wave
functions and to study their mass splittings.

So, we shall solve the Schr\" odinger equation with the linear
``hyper-radial" potential, see Appendix \ref{s:Hrad2}, use its
solutions as unperturbed states, and treat the second term in Eq.
(\ref{e:exp1}) as the (lowest-order) perturbation in
``hyper-spherical harmonics", see Sects. \ref{s:HSY-D},
\ref{s:summary} and Appendix \ref{s:HSang}.

\subsection{Non-perturbative approximations}
\label{s:nonpert}

The hyper-spherical formalism is often advertised as an economical
way to tackle the general three-body problems in atomic and
nuclear physics, but, in fact, it seems as if it had been
tailor-made for the Y-string potential. We refer to specialized
papers for technical aspects of this method (see for instance
Refs. \cite{Krivec}). Here we recall just what is needed for our
purposes; then, we use it below to tackle the problem
non-perturbatively.

\subsubsection{Y-string problem in hyper-spherical coordinates}
\label{app:HS}

We expand the three-quark wave function in hyper-spherical
coordinates as
\begin{eqnarray}
\Psi(R,\chi,\hat{\rho},\hat{\lambda}) & = & \sum_c \psi_c(R) {\cal
Y}_c(\chi,\hat{\rho},\hat{\lambda})
\end{eqnarray}
where $c = l_1,l_2,L,m$ and $K=l_1 + l_2 + 2m$ (occasionally one
uses $[K]$ to denote the complete set of hyperspherical quantum
numbers  $K,l_1,l_2,L,m)$ and
\begin{eqnarray}
{\cal
Y}_c(\chi,\hat{\rho}=\Omega_{\rho},\hat{\lambda}=\Omega_{\lambda})
& = & u_{l_1,l_2,m}(\chi)
[Y_{l_1}(\hat{\rho}) \otimes Y_{l_2}(\hat{\lambda})]_{L} \\
u_{l_1,l_2,m}(\chi) & = & N_{l_1,l_2,m} F(-m,l_1+l_2+m+2|l_2 +
\frac{3}{2}|\sin^2\chi) (\cos\chi)^{l_1}(\sin\chi)^{l_2} \\
N_{l_1,l_2,m} & = & \frac{1}{\Gamma(l_2 + 3/2)}
\sqrt{\frac{2(2m+l_1+l_2 +2)\Gamma(m+l_1+l_2+2)\Gamma(m+l_2+3/2)}
{m!\Gamma(m+l_1+3/2)}}
\end{eqnarray}
where $F(-n,n+a|c|z)$ is the Jacobi function
\begin{eqnarray}
F(-n,n+a|c|z) & = & \sum_{s=0}^{n}(-1)^s
\frac{n!\Gamma(c)\Gamma(a+n+s)}{(n-s)! s!
\Gamma(a+n)\Gamma(c+s)}z^s
\end{eqnarray}
and $\Gamma(c)$ is Euler's Gamma function.

An important property of the hyper-spherical formalism is that a
complicated three-body problem reduces to a set of (coupled)
differential equations involving only the hyper-radial matrix
elements of the potential $V(R)_{[K][K^{'}]}= \langle
\Psi(\chi,\theta)_{[K^{'}]}| V(R,\chi, \theta) |
\Psi(\chi,\theta)_{[K]} \rangle_{\rm hyp-ang.}$. The Schr\"
odinger equation of the three-quark system then becomes a set of
infinitely many coupled equations,
\begin{eqnarray}
- \frac{1}{2\mu}\left[\frac{d^2}{dR^2} + \frac{5}{R}\frac{d}{dR} -
\frac{K(K+4)}{R^2} + 2\mu E \right]\psi_c(R) +
\sqrt{\frac{3}{2}}\sigma R \sum_{c'} C_{c,c'}\psi_{c'}(R) = 0
\end{eqnarray}
with
\begin{eqnarray}
C_{c',c} &=& \langle \Psi(\chi,\theta)_{c'}| \sqrt{1 + \sin 2 \chi
|\sin \theta_{12}|} | \Psi(\chi,\theta)_{c} \rangle_{\rm hyp-ang.}
\nonumber \\
& = & \int_0^{\pi/2} \cos\chi^2 \sin\chi^2 d \chi \int
d\Omega_{\rho} d\Omega_{\lambda} {\cal Y}_{c'}^* \sqrt{1 + \sin
2\chi |\sin\theta_{12}|}{\cal Y}_c
\end{eqnarray}
where $\cos\theta_{12} = \hat{\rho}\cdot\hat{\lambda}$.

In special cases, such as the present one, some (finite) subset(s)
of equations may decouple, due to the symmetries of the
interaction potential. The spectrum of the system is then reduced
to finding the eigenvalues of the following differential equation
\begin{eqnarray}
\left[\frac{d^2}{dR^2} + \frac{5}{R}\frac{d}{dR} -
\frac{K(K+4)}{R^2} - a_c R + k^2 \right] \psi_c(R) = 0
\end{eqnarray}
where
\begin{eqnarray}
a_c & =& 2 \mu \sqrt{\frac{3}{2}}\sigma C_{c,c}\\
k^2 & = & 2 \mu E
\end{eqnarray}
and the coupling matrix $C_{c'c}$
\begin{eqnarray}
C_{c',c} &=& \sum_{L'} X_{c',c,L'} \int_0^{\pi/2}
u_{c'}^{*}(\chi)u_{c}(\chi)\,\cos^2\chi\sin^2\chi d\chi
\int_{-1}^{1} d\cos\theta_{12} P_{L'}(\cos\theta_{12}) \sqrt{1 +
\sin 2\chi\,|\sin\theta_{12}|}
\end{eqnarray}
can be evaluated using the by now familiar angular re-coupling
matrix $X_{c',c,L'}$,
\begin{eqnarray}
X_{c',c,L'} & = &
\sqrt{(2l_{1'}+1)(2l_{2'}+1)(2l_{1}+1)(2l_{2}+1)}/2
 \nonumber \\
& \times & (-1)^{L'+l_{1'}+l_{2'}}
W(l_{1},L',L,l_{2'};l_{1'},l_{2})
(l_{1'},0,l_{1},0|L',0)(l_{2'},0,l_{2},0|L',0)
\end{eqnarray}
and the following two-dimensional integrals
\begin{eqnarray}
&& \int_{-1}^{1} P_{L}(\,\cos\theta) \,d\cos\theta \int
_{0}^{\frac{\pi}{2}} |u_{c}(\chi)|^{2}\, {\sqrt{1 + |\sin\theta|
\sin\,2\chi}} \,{\left( \cos\chi\,\sin\chi \right) }^2\, d\chi \
\end{eqnarray}
We may numerically evaluate the necessary integrals, see Appendix
\ref{s:NPYstring}, so as to evaluate the hyper-angular matrix
elements and recast the Schr\" odinger equation with the exact
potential $V_{\rm Y}$ into a set of (coupled) hyper-radial
equations for each ``hyper-spherical harmonic" that can be solved
numerically, see Appendix \ref{s:Hrad}.

\subsubsection{The complete string potential in hyper-spherical coordinates}
\label{app:HS}

\begin{figure}[tbp]
\centerline{\includegraphics[width=4.5in,,keepaspectratio]{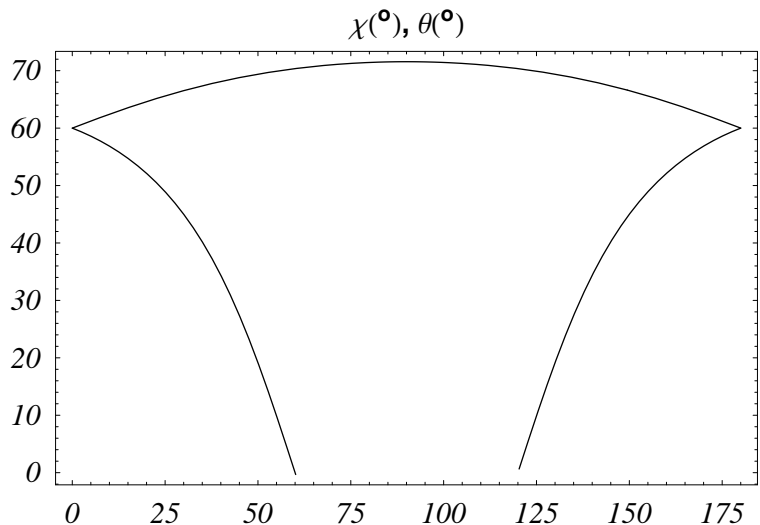}}
\caption{The boundary in the $\chi$ vs. $\theta$ plane, between
the regions in which the two- and the three-string potentials are
appropriate, see Eqs. (\ref{e:boundary}).}
\label{f:bound1}
\end{figure}

To finally solve the exact string problem, one must relax the
Y-string potential approximation Eq. (\ref{hypVY1a}) and use the
exact potential, i.e., all four of its incarnations, Eqs.
(\ref{hypVY1a},\ref{hypVY1b},\ref{hypVY1c},\ref{hypVY1d})
depending on the angles formed by the three quarks. This means
that one must first determine the boundary in the $\chi$ vs.
$\theta$ plane between the regions in which the two- and the
three-string potentials are appropriate, see Eqs.
(\ref{e:boundary}). There are three such boundaries, determined by
the three inequalities, that merge continuously one into another
at two ``contact points", see Fig. \ref{f:bound1}.
\begin{eqnarray}
\cot\chi_1(\theta) &=& \frac{-1}{{\sqrt{3}}\,\cos\theta +
\sin\theta}, \nonumber \\
\cot\chi_2(\theta) &=& \frac{1}{{\sqrt{3}}\,\cos\theta -
\sin\theta},  \nonumber \\
\cot\chi_3(\theta) &=& \frac{1}{3}{\sqrt{5 - 2\,{\cos}^2\theta -
2\,|\sin\theta|\,{\sqrt{4 - \cos^2\theta}}}}. \label{e:boundary} \
\end{eqnarray}
The hyper-angular matrix elements with these boundary conditions
are evaluated in Appendix \ref{s:NPCompletestring}. The rest of
the calculation proceeds as in the case of the Y-string.

\section{Results}
\label{s:results}

\subsection{The first-order perturbation theory}
\label{s:pert-result}

\subsubsection{Y-string three-quark potential expanded to first powers of
hyper-angles} \label{s:HSY-D}

In Appendix \ref{app:HS} we have calculated the  matrix elements
$\langle \,{\rm N}_{\rm K}, {\rm K} \, [SU_{\rm FS}(6),L^P] \, |\,
\frac{2\,|{\bm \rho} \times {\bm \lambda}|}{R} \, |\,{\rm N}_{\rm
K}, {\rm K} \, [SU_{\rm FS}(6),L^P] \, \rangle$ for the
states/$SU_{\rm FS}(6)$ multiplets of the three lowest ${\rm
N}_{\rm K}$(=0,1,2) bands: $|\,{\rm N}_{\rm K}=0, {\rm K}=0,
[56,0^{+}] \, \rangle$, $|\,{\rm N}_{\rm K}=1, {\rm K}=1,
[70,1^{-}] \, \rangle$, and $|\,{\rm N}_{\rm K}=1, {\rm K}=0,
[56^{\prime},0^{+}] \, \rangle$, $|\,{\rm N}_{\rm K}=2, {\rm K}=2,
[56,2^{+}] \, \rangle$, $|\,{\rm N}_{\rm K}=2, {\rm K}=2,
[70,0^{+}] \, \rangle$, $|\,{\rm N}_{\rm K}=2, {\rm K}=2,
[70,2^{+}] \, \rangle$, $|\,{\rm N}_{\rm K}=2, {\rm K}=2,
[20,1^{+}] \, \rangle$ where we separate each matrix element into
its hyper-radial and hyper-angular parts:
\begin{eqnarray}
\langle \, \psi_{{\rm N}_{\rm K}, [{\rm K}]} \, |\, \frac{2\,|{\bm
\rho} \times {\bm \lambda}|}{R} \, |\, \psi_{{\rm N}_{\rm K},
[{\rm K}]} \, \rangle &=& \langle \, \psi_{{\rm N}_{\rm K}, {\rm
K}} \, | \, R \, |\, \psi_{{\rm N}_{\rm K}, {\rm K}} \,
\rangle_{\rm hyp-rad} \langle \, \psi_{[{\rm K}]} \, |\,\sin2\chi
\, \sin\theta_{\rho\lambda} \, |\, \psi_{[{\rm K}]} \,
\rangle_{\rm hyp-ang}, \
\end{eqnarray}
where $[{\rm K}]$ denotes all the quantum numbers, such as the $L,
l_{\rho}, l_{\lambda}$ and their magnetic quantum numbers,
associated with ${\rm K}$. The hyper-radial matrix element can be
calculated using the virial theorem, see Appendix \ref{s:Hrad1} ,
as follows:
\begin{eqnarray}
\sigma \sqrt{\frac32} \langle \, \psi_{{\rm N}_{\rm K}, {\rm K}}
\, | \, R \, |\, \psi_{{\rm N}_{\rm K}, {\rm K}} \, \rangle_{\rm
hyp-rad} &=& \frac{2}{3} \langle \, \psi_{{\rm N}_{\rm K}, {\rm
K}} \, | \, H \, |\, \psi_{{\rm N}_{\rm K}, {\rm K}} \,
\rangle_{\rm hyp-rad} = \frac{2}{3} E_{{\rm N}_{\rm K}, {\rm K}}.
\
\end{eqnarray}
The calculated energies of states with various values of K and L
are listed in Table \ref{tab:drcc2n}.
\begin{table}[tbh]
\begin{center}
\caption{The values of the unperturbed energy $E_{{\rm N}_{\rm
K},{\rm K}}$, the total energy $E_{\rm K}+\delta E_{\rm K,L}$ (in
units of $\left(\sqrt{\frac32} \frac{{\sigma}\, {\hslash}}{\sqrt{2
\mu}}\right)^{\frac{2}{3}}$) and the three-body potential matrix
elements' hyper-angular perturbation $\langle \frac{2\,|{\bm \rho}
\times {\bm \lambda}|}{R^2}\rangle_{\rm hyp-ang.} = \langle
\sin2\chi \,|\sin \theta| \rangle_{\rm hyp-ang.}$ as well as some
intermediate steps, for the various K = 0,1,2 states (with all
allowed orbital waves L).}
\begin{tabular}{cccccccc}
\hline \hline K & ${\rm N}_{\rm K}$ & $E_{{\rm N}_{\rm K},{\rm
K}}^{(0)}$ & $[SU(6),L^P]$ & $\frac{1}{3}\langle \sin2\chi \,|\sin
\theta| \rangle$ & $E_{{\rm N}_{\rm K},{\rm K}}^{(0)}+\delta
E_{{\rm N}_{\rm K},{\rm K},{\rm L}}$ & = $E_{{\rm N}_{\rm K},{\rm K},{\rm L}}$\\
\hline
0 & 0 & 3.8175 & $[56,0^+]$ & $\frac29$ & $\frac{11}{9}$ 3.8175 & 4.6658 \\
1 & 0 & 4.6582 & $[70,1^-]$ & $\frac{2}{9}$ & $\frac{11}{9}$ 4.6582 & 5.6934 \\
0 & 1 & 5.2630 & $[56,0^+]$ & $\frac{2}{9}$ & $\frac{11}{9}$ 5.2630 & 6.4326 \\
\hline
2 & 0 & 5.4290 & $[70,0^+]$ & $\frac{8}{45}$ & $\frac{53}{45}$ 5.4290 & 6.3942 \\
2 & 0 & 5.4290 & $[56,2^+]$ & $\frac{44}{225}$ & $\frac{269}{225}$ 5.4290  & 6.4907 \\
2 & 0 & 5.4290 & $[70,2^+]$ & $\frac{52}{225}$ & $\frac{277}{225}$ 5.4290  & 6.6837 \\
2 & 0 & 5.4290 & $[20,1^+]$ & $\frac{4}{15}$ & $\frac{19}{15}$ 5.4290 & 6.8767 \\
\hline
\end{tabular}
\label{tab:drcc2n}
\end{center}
\end{table}
Note that the hyper-radial matrix elements of the linear
hyper-radial potential are identical for all K=2 band hyper-radial
(ground) states. Therefore, all the energy differences among
various K=2 multiplets are integer multiples of the energy
splitting ``unit" $\Delta_{Y} = \frac{2}{75} \sigma \sqrt{\frac32}
\langle R \rangle_{{\rm N},{\rm K}} = 0.0327 \sigma \langle R
\rangle_{{\rm N},{\rm K}} = \frac{8}{225} E_{{\rm N}_{\rm K}, {\rm
K}}$, which is just another manifestation of the validity of the
Bowler-Tynemouth theorem \cite{BT83} for this three-body potential
solved in hyper-spherical coordinates. 

This theorem has already been confirmed by J.M. Richard and P.
Taxil \cite{RT90}, in the hyper-spherical formalism with linear
two-body potentials. Our new contribution here is the (first)
proof of this theorem, with the ``area term" three-body potential
as a perturbation, that holds even with non-harmonic oscillator
unperturbed states. As in the case of Bowler and Tynemouth
\cite{BT83} the crucial ingredient of the proof is the permutation
symmetry of the three-body potential. Other simple three body
potentials that lack this permutation symmetry do not conform to
this theorem.

Returning to the details of the spectrum, note, however, that the
Roper $[56,0^+]$ multiplet (6.433) is roughly half-way between the
$[70,0^+]$ (6.394) and the $[56,2^+]$ (6.491) multiplets, and for
most practical intents and purposes degenerate with them both, as
their mass difference is just one $\Delta_Y$. The ratios of the
K=2 and other K-values hyper-radial matrix elements need not be
integers, or rational numbers any more, except in special cases,
like in the harmonic oscillator, or in the Coulomb potential. In
the Y-string potential the upward shift of the Roper $[56,0^+]$
(5.263 $\to$ 6.433) is proportional to its unperturbed energy
(4.082), as is the upward shift of the K=1 odd-parity states
(4.658 $\to$ 5.693), and the coefficient of proportionality
$\frac{11}{9}$ is equal for these two states.

In other words, the Y-string potential does not move the Roper
below the K=1 odd-parity resonances, at least not in 
perturbation theory. Note, however, that the largest correction in
this hyper-spherical first-order perturbative approximation
relative to the unperturbed value is 27\%, which still does not
justify a perturbative treatment. We shall therefore try and
estimate the effects of the exact potential, i.e. the whole power
series at once.

\subsection{Non-perturbative results}
\label{s:allorder-result}

\subsubsection{The three-string potential}
\label{s:HSY-D1}

Instead of expanding $\sqrt{1 + x}$, where $x=\sin\theta
\sin\,2\chi$, in a power series we may numerically integrate the
double integral
\begin{eqnarray}
\int _{0}^{\pi } P_{L}(\,\cos\theta) \,\sin\theta \,d\theta \int
_{0}^{\frac{\pi}{2}} |u_{c,K,L}(\chi)|^{2}\, {\sqrt{1 + \sin\theta
\sin\,2\chi}} \,{\left( \cos\chi\,\sin\chi \right) }^2\, d\chi .\
\end{eqnarray}
Such hyper-angular potential matrix elements are the coefficients
multiplying the linear hyper-radial potential that appears in the
(new) hyper-radial Schr\" odinger equation; that equation, in
turn, can be solved exactly, i.e. without the use of the
perturbation theory, and the resulting energy eigenvalues are
listed in Table \ref{tab:drcc2n1}.
\begin{table}[tbh]
\begin{center}
\caption{The values of the unperturbed energy $E_{{\rm N}_{\rm
K},{\rm K}}^{(0)}$, the total energy $E_{\rm K}$ (in units of
$\left(\sqrt{\frac32} \frac{{\sigma}\, {\hslash}}{\sqrt{2
\mu}}\right)^{\frac{2}{3}}$) and the three-body potential matrix
elements' hyper-angular exact non-perturbative matrix element
$\langle V_{\rm Y} \rangle_{\rm hyp-ang.} = \langle \sqrt{1 + \sin
2 \chi |\sin \theta|} \rangle_{\rm hyp-ang.}$, for the various K =
0,1,2 states (with all allowed orbital waves L).}
\begin{tabular}{cccccccc}
\hline \hline K & ${\rm N}_{\rm K}$ & $E_{{\rm N}_{\rm K},{\rm
K}}^{(0)}$ & $[SU(6),L^P]$ & $\langle \sqrt{1 + \sin 2 \chi |\sin
\theta|} \rangle_{\rm hyp-ang.}$ & $\langle V_{\rm Y}
\rangle_{h.a.}^{2/3} E_{\rm K}^{(0)}$ & = $E_{{\rm N}_{\rm K},{\rm K},{\rm L}}$\\
\hline
0 & 0 & 3.8175 & $[56,0^+]$ & 1.2876 & 1.18355$\times$3.8175 & 4.5182 \\
1 & 0 & 4.6582 & $[70,1^-]$ & 1.2876 & 1.18355$\times$4.6582 & 5.5132 \\
0 & 1 & 5.2630 & $[56,0^+]$ & 1.2876 & 1.18355$\times$5.2630 & 6.2290 \\
\hline
2 & 0 & 5.4290 & $[70,0^+]$ & 1.2350 & 1.15110$\times$5.4290 & 6.2493 \\
2 & 0 & 5.4290 & $[56,2^+]$ & 1.2560 & 1.16411$\times$5.4290  & 6.3199 \\
2 & 0 & 5.4290 & $[70,2^+]$ & 1.2981 & 1.18998$\times$5.4290  & 6.4604 \\
2 & 0 & 5.4290 & $[20,1^+]$ & 1.3402 & 1.21557$\times$5.4290 & 6.5993 \\
\hline
\end{tabular}
\label{tab:drcc2n1}
\end{center}
\end{table}
We can compare these exact results with the first-order
perturbation theory: the energy eigenvalues, after rescaling, see
Appendix \ref{s:Hrad2}, are proportional to $E \sim
\left(\sqrt{\frac32} \frac{{\sigma}\, {\hslash}}{\sqrt{2
\mu}}\langle \sqrt{1 + \sin 2 \chi |\sin \theta|} \rangle_{\rm
hyp-ang.}\right)^{\frac{2}{3}}$ (modulo an overall hyper-radial
dependent factor that is the same for all K=2 ground states). This
can be (double) Taylor-expanded and yields $\left(\sqrt{\frac32}
\frac{{\sigma}\, {\hslash}}{\sqrt{2 \mu}}\left(1 + \frac12 \langle
\sin 2 \chi |\sin \theta| \rangle_{\rm hyp-ang.}
\right)\right)^{\frac{2}{3}} \simeq \left(\sqrt{\frac32}
\frac{{\sigma}\, {\hslash}}{\sqrt{2
\mu}}\right)^{\frac{2}{3}}\left(1 + \frac{1}{3} \langle \sin 2
\chi |\sin \theta| \rangle_{\rm hyp-ang.} \right)$, which is the
first perturbative correction.

Thus we may expect the exact result to be smaller than the
first-order perturbative result on two accounts: 1) from the
inclusion of all orders in the expansion of the square-root in the
potential: the even- and odd-order terms have opposite signs (this
is an alternating series), see e.g. the first ten terms
\begin{eqnarray}
\sqrt{1 + x} &=& 1 + \frac{x}{2} - \frac{x^2}{8} + \frac{x^3}{16}
- \frac{5\,x^4}{128} + \frac{7\,x^5}{256} - \frac{21\,x^6}{1024} +
\frac{33\,x^7}{2048} - \frac{429\,x^8}{32768} +
\frac{715\,x^9}{65536} + {\cal O}(x^{10}), \
\end{eqnarray}
with rapidly decaying coefficients, meaning that the most
important correction to the linear $\frac{x}{2}$ term is the
negative $\frac{x^2}{8}$ one; and 2) from the inclusion of all
orders in the expansion of the two-thirds-root. Manifestly, both
of these effects lead to the breaking of the ``integer-value rule"
of Bowler theorem for the energy splittings of the K=2 multiplets,
but only at the 1\% level.

Note the overall reduction of the potential matrix elements, as
compared with the first-order perturbation theory results: the
largest correction in the hyper-spherical non-perturbative
approach relative to the unperturbed value is 22\%, which is less
than the 27\% in the hyper-spherical non-perturbative
approximation. This fact justifies {\it ex post facto} the
perturbative treatment in Sect. \ref{s:pert-result}.


\subsubsection{The complete string results}
\label{s:2string-result}

Finally, the results of the solution to the complete string
potential problem are shown in Table \ref{tab:drcc2n2}, which
shows a slight ${\cal O}(< 1\%)$, yet clear increase, across the
board, of the total energy over the Y-string approximation results
in Table \ref{tab:drcc2n1}.

The overall shifts of energies in this calculation would hardly
justify the effort it took to complete it, were it not for their
(relative) effects on the K=2 level splittings, which are
substantial: The violations of the ``Bowler-Tynemouth theorem"
increase to 13\% and 9\% for the $[20,1^+]$-$[70,2^+]$, and
$[70,2^+]$-$[56,2^+]$ mass differences, respectively, as compared
with the $[56,2^+]$-$[70,0^+]$ one.
\begin{table}[tbh]
\begin{center}
\caption{The values of the unperturbed energy $E_{{\rm N}_{\rm
K},{\rm K}}^{(0)}$, the total energy $E_{\rm K}$ (in units of
$\left(\sqrt{\frac32} \frac{{\sigma}\, {\hslash}}{\sqrt{2
\mu}}\right)^{\frac{2}{3}}$) and the exact three-body potential
hyper-angular non-perturbative matrix element $\langle V_{\rm
string} \rangle_{\rm hyp-ang.}$, taken from Table
\ref{tab:drcc2m2}, as well as some intermediate steps, for the
various K = 0,1,2 states (with all allowed orbital waves L).}
\begin{tabular}{cccccccc}
\hline \hline K & ${\rm N}_{\rm K}$ & $E_{{\rm N}_{\rm K},{\rm
K}}^{(0)}$ & $[SU(6),L^P]$ & $\langle V_{\rm string} \rangle_{\rm
hyp-ang.}$ & $\langle V_{\rm string}
\rangle^{2/3} E_{\rm K}^{(0)}$ & = $E_{{\rm N}_{\rm K},{\rm K},{\rm L}}$\\
\hline
0 & 0 & 3.8175 & $[56,0^+]$ & 1.2891 & 1.18449$\times$3.8175 & 4.5218 \\
1 & 0 & 4.6582 & $[70,1^-]$ & 1.2891 & 1.18449$\times$4.6582 & 5.5176 \\
0 & 1 & 5.2630 & $[56,0^+]$ & 1.2891 & 1.18449$\times$5.2630 & 6.2340 \\
\hline
2 & 0 & 5.4290 & $[70,0^+]$ & 1.2401 & 1.15943$\times$5.4290 & 6.2665 \\
2 & 0 & 5.4290 & $[56,2^+]$ & 1.2584 & 1.16886$\times$5.4290 & 6.3279 \\
2 & 0 & 5.4290 & $[70,2^+]$ & 1.2985 & 1.19022$\times$5.4290 & 6.4617 \\
2 & 0 & 5.4290 & $[20,1^+]$ & 1.3404 & 1.21567$\times$5.4290 & 6.5999 \\
\hline
\end{tabular}
\label{tab:drcc2n2}
\end{center}
\end{table}
Note, that the Roper multiplet $[56,0^+]$ (6.2340) has moved below
the $[70,0^+]$ (6.2665) multiplet, but remains, for all practical
purposes degenerate with it.

\section{Summary and Discussion}
\label{s:summary}

\begin{figure}[tbp]
\centerline{\includegraphics[width=4.5in,,keepaspectratio]{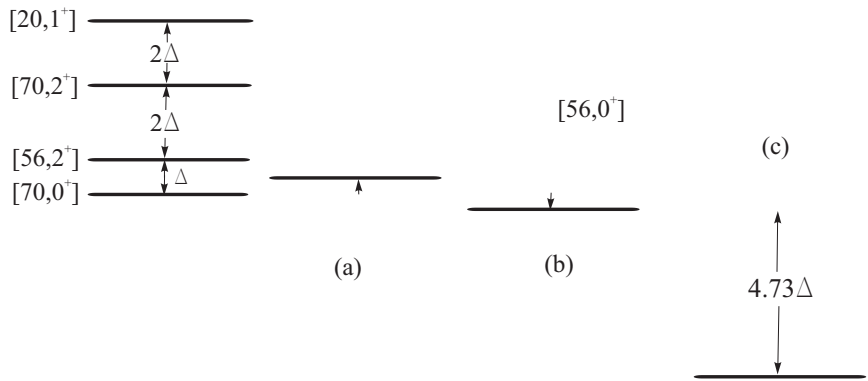}}
\caption{Depiction of the energy splitting of the K = 2 states of
the hyper-spherical linear potential spectrum due to attractive
three-body potentials: (a) the first-order perturbation
approximation to the first-power expansion of the ``three-string"
potential (see text for definition); (b) exact non-perturbative
results of the ``complete-string" potential and (c) the (rescaled)
$\Delta$-string potential results of J.M. Richard and Taxil
\cite{RT90}. The left-hand side of the diagram involving the
$[20,1^+], [70,2^+], [56,2^+], [70,0^+]$ multiplets is common to
both kinds of potentials, follows the Bowler-Tynemouth rule 2:2:1
to 1\%; only the position of the $[56,0^+]$ multiplet (containing
the Roper resonance) is variable.} \label{spectrum}
\end{figure}

In summary, we have studied the low-lying states in the three
quark spectra confined by a pure Y-string potential, i.e. without
any two-quark potentials, in three different approximations, see
Table \ref{T:Finalspectrum} and Fig. \ref{spectrum}.

An attractive Y-string potential splits the N=K=2 band states into
degenerate SU$_{FS}$(6) multiplets: $[20,1^{+}], [70,2^{+}],
[56,2^{+}], [70,0^{+}]$, ordered in descending mass, and following
approximately the separation rule of 2:2:1. The best accuracy
results lead to 2.25:2.18:1 splitting i.e. to a maximum violation
of this rule is less than $13\%$.

The mass difference between the first (hyper-) radial excitation
of the ground state, that is the ``Roper multiplet"
$[56^{'},0^+]$, and the odd-parity K=N=1 $ [70,1^{-}]$ multiplet
is entirely determined by the difference between and the first
(hyper-) angular and the first (hyper-) radial excitation
eigen-energies in a linearly rising hyper-radial potential, which
is always negative. In other words, the Roper resonance cannot be
lowered below the odd-parity K=N=1 states in this potential,
irrespective of the string tension constant and the quark masses,
which are the only free parameters.

\begin{table}[tbh]
\begin{center}
\caption{The eigen-energies (in units of $\left(\sqrt{\frac32}
\frac{{\sigma}\, {\hslash}}{\sqrt{2 \mu}}\right)^{\frac{2}{3}}$)
of the unperturbed solution to the hyper-central approximation,
(see the text) $E_{{\rm N}_{\rm K},{\rm K}}^{(0)}$, one
perturbative ($E_{{\rm N}_{\rm K},{\rm K},{\rm L}}^{(1)}$), and
two non-perturbative approximations ($E_{{\rm N}_{\rm K},{\rm
K},{\rm L}}^{(2)}, ~E_{{\rm N}_{\rm K},{\rm K},{\rm L}}^{(3)})$ to
the Y-string potential, where the fourth one $E_{\rm K}^{(Y)}$ is
the exact (numerical) result, for the various low-lying K = 0,1,2
states (with all allowed orbital waves L). The last column shows
the results for the $\Delta$-string potential of J.M. Richard and
Taxil \cite{RT90}.}
\begin{tabular}{cccccccccc}
\hline \hline K & ${\rm N}_{\rm K}$ & $[SU(6),{\rm L^P}]$ &
$E_{{\rm N}_{\rm K},{\rm K}}^{(0)}$ & $E_{{\rm N}_{\rm K},{\rm
K},{\rm L}}^{(1)}$ & $E_{{\rm N}_{\rm K},{\rm K},{\rm L}}^{(2)}$ &
$E_{{\rm N}_{\rm K},{\rm K},{\rm L}}^{(Y)}$ &
$E_{{\rm N}_{\rm K},{\rm K},{\rm L}}^{(\Delta)}$ & \\
\hline
0 & 0 & $[56,0^+]$ & 3.8175 & 4.6658 & 4.5182 & 4.5218 & 5.3592 \\ 
1 & 0 & $[70,1^-]$ & 4.6582 & 5.6934 & 5.5132 & 5.5176 & 6.5395 \\ 
0 & 1 & $[56,0^+]$ & 5.2630 & 6.4326 & 6.2290 & 6.2340 & 7.3885 \\ 
\hline
2 & 0 & $[70,0^+]$ & 5.4290 & 6.3942 & 6.2493 & 6.2665 & 7.5409 \\ 
2 & 0 & $[56,2^+]$ & 5.4290 & 6.4907 & 6.3199 & 6.3279 & 7.5731 \\ 
2 & 0 & $[70,2^+]$ & 5.4290 & 6.6837 & 6.4604 & 6.4617 & 7.6377 \\ 
2 & 0 & $[20,1^+]$ & 5.4290 & 6.8767 & 6.5993 & 6.5999 & 7.7022 \\ 
\hline
\end{tabular}
\label{T:Finalspectrum}
\end{center}
\end{table}

Next we compare our Y-string three-quark potential results with
the $\Delta$-shaped string ones, as obtained in Ref. \cite{RT90}.
A quick look at Table \ref{T:Finalspectrum} shows that the
$\Delta$-string eigen-energies are substantially higher than the
corresponding Y-string ones, with identical string tension
$\sigma$.

In order to check the (re)scaling hypothesis {\it viz.} if the
low-lying spectra are identical after a (re)scaling of the string
tension $\sigma_Y \to f \sigma_{\Delta}$, we tabulate the rescaled
$\Delta$-string energies next to the exact Y-string ones in Table
\ref{T:comparespectrum}. The string tension scaling factor $f$ is
fixed at $\frac{5\,\pi}{32}\sqrt{\frac{3}{2}} 1.2891 = 0.775001$,
which ensures that the three lowest-lying states/bands, {\it viz.}
$(K=0,N_K=0,1)$ and $(K=1,N_K=0)$, have identical eigen-energies,
see Table \ref{T:comparespectrum}. Note that this value of $f$ is
substantially different from $0.5493$, which was used in Ref.
\cite{Bonn01}. The energy rescaling factor is just the $\frac23$
power of $f$: $f^{2/3} = \left(\frac{5\,\pi}{32}\sqrt{\frac{3}{2}}
1.2891\right)^{\frac{2}{3}} = 0.843726$.
\begin{table}[tbh]
\begin{center}
\caption{The eigen-energies (in units of $\left(\frac{{\sigma}\,
{\hslash}}{\sqrt{2 \mu}}\right)^{\frac{2}{3}}$) of the Y-string
potential $E_{\rm K}^{(Y)}$ and the $\Delta$-string potential
results of J.M. Richard and Taxil \cite{RT90} rescaled by factor
$f^{2/3} = 0.843726$, see text, for the low-lying K = 0,1,2
states.}
\begin{tabular}{cccccc}
\hline \hline K & ${\rm N}_{\rm K}$ & $[SU(6),{\rm L^P}]$ &
$E_{{\rm N}_{\rm K},{\rm K},{\rm L}}^{(Y)}$ &
$f^{2/3} E_{{\rm N}_{\rm K},{\rm K},{\rm L}}^{(\Delta)}$\\
\hline
0 & 0 & $[56,0^+]$ & 5.1761 & 5.1761 \\
1 & 0 & $[70,1^-]$ & 6.3160 & 6.3160 \\
0 & 1 & $[56,0^+]$ & 7.1360 & 7.1360 \\
\hline
2 & 0 & $[70,0^+]$ & 7.1733 & 7.2832 \\
2 & 0 & $[56,2^+]$ & 7.2437 & 7.3143 \\
2 & 0 & $[70,2^+]$ & 7.3968 & 7.3767 \\
2 & 0 & $[20,1^+]$ & 7.5550 & 7.4390 \\
\hline
\end{tabular}
\label{T:comparespectrum}
\end{center}
\end{table}
One can see that the overall effect of the Y-string potential on
the spectrum does not amount to a mere rescaling of the energy,
i.e. of the string tension $\sigma$ in the $\Delta$-string
potential: the Y-string potential pushes all of the K=2 states up
in energy, some less others more, whereas the $\Delta$-string
shifts some K=2 states up, others down and leaves third states
unchanged relative to the Y-string states. Moreover, the size of
the energy splitting of the K=2 band states in the Y-string
(0.381684) is more than two times bigger than in the rescaled
$\Delta$-string potential (0.155752).

Of course, the ordering of the shells/bands and of states within
shells remains identical in both potentials, because it is
controlled by the permutation symmetry of the states and of the
potentials, as well as by the convexity of the potential in the
hyper-radial coordinate. The only qualitative difference between
the $\Delta$- and Y-string potentials that we found is that the
``Roper" lies lower, as compared with the lowest lying K=2 state,
in the $\Delta$-string than in the Y-string potential.

So we may conclude that, neglecting relativistic and (the very
important) HFI effects, the three lowest-lying bands of states
that form the (only) set of well-established resonances so far, do
{\it not} allow a clear distinction to be made between these two
types of potentials. This may, but need not, be a surprise, as
these two string potentials have (very) different functional forms
in configuration space, which we (naively) expected to predict
different physics. But it turns out that the Bowler-Tynemouth
theorem holds for general hyper-radial potentials, not just for
the harmonic oscillator one, and then the separation of states in
the K=2 shell is (tightly) constrained by the permutation
symmetry.

The technical complexities of the Y-string potential, such as the
orbital angular momentum dependence, which is responsible for some
basic features of the spectrum, and the coupling of the radial and
angular motions, encountered above are difficult to deal with
accurately with methods that are not based on the hyper-spherical
coordinates. The latter technique is not widely familiar to many
practitioners in this field, however. The hyper-spherical
formalism has one unfortunate disadvantage: the relativistic
kinetic energy of three quarks is not a function of only the
hyper-radius and as such the method does not (readily) extend to
relativistic energies. For this reason we expect the present
results to be relevant and applicable only to heavy-quark baryon
spectroscopy, such as $ccc$.

That may explain the reason why this three-string potential has
often been replaced in practical calculations by various two-body
approximations, that are more easily handled by standard methods.
Several two-body approximations to this potential have been
devised and compared with the exact Y-string potential in various
limits \cite{fabr97,Bonn01,Grenoble03}. The results of the two
most systematic papers so far \cite{fabr97,Grenoble03} are
inconclusive, however, because the former did not calculate the
spectrum and the latter is based on the hyper-radial
approximation, which does not split the various SU(6) multiplets
within the K=2 shell.

There is only one possible clue to the shape of the confining
potential in the lower end of the baryon resonance spectrum, {\it
viz.} the Roper resonances (abnormally low) mass, that perhaps
could be used to draw conclusions about the existence and/or
preponderance of one kind of potential over the other. We have
shown, however, that the Y-shaped string always leads to a Roper
resonance that is heavier than the lowest-lying odd-parity
resonance, just like the $\Delta$-string.

This does not mean, however, that there need not be any
differences between the spectra of the Y- and the $\Delta$-string
potentials, rather, it means that one must go to the higher lying
bands of states, and in particular to higher orbital angular
momentum states, in order to see these differences. It remains to
be seen just how high is high enough.

\section*{Acknowledgments}
\label{ack}

One of us (V.D.) wishes to thank Prof. S. Fajfer for her
hospitality at the Institute Jo\v zef Stefan, Ljubljana, where
this work was started and to Profs. H. Toki and A. Hosaka for
their hospitality at RCNP, Osaka University, where it was
continued. V.D. also thanks Dr. Rajmund Krivec for providing him
with Ref. \cite{Krivec}.

\appendix

\section{The hyper-spherical approach}
\label{s:Y-HSpert}

\subsection{Brief review}
\label{s:HSreview}

The matrix elements/expectation values of the three-body potential
$V$ between the hyper-spherical harmonics ${\cal
Y}_{K,L,m_{i}}(\Omega_{5})$ and ${\cal
Y}_{K^{'},L^{'},m_{i}^{'}}(\Omega_{5})$ are defined by
\begin{eqnarray}
\langle V_{\rm Y}(R)\rangle_{[K][K^{'}]} &=& \int~ {\cal
Y}_{K,L,m_{i}}^{*}(\Omega_{5}) \, V({\bm \rho }, {\bm \lambda })\,
{\cal Y}_{K^{'},L^{'},m_{i}^{'}}(\Omega_{5}) \, d\Omega ^{(5)}
\nonumber \\
&=& \int_{0}^{\pi/2} \cos ^{2}\chi \, \sin^{2}\chi \,d\chi \,
\int_{0}^{\pi}\,\sin \theta \,d\theta \,{\cal
Y}_{K,L,m_{i}}^{*}(\Omega_{5}) \, V({\bm \rho }, {\bm \lambda })\,
{\cal Y}_{K^{'},L^{'},m_{i}^{'}}(\Omega_{5}) . \
\label{vhpc}
\end{eqnarray}
where
\begin{eqnarray}
\cos\theta &=& \cos\theta_{\rho} \, \cos\theta_{\lambda} +
\cos(\varphi_{\rho} - \varphi_{\lambda})\, \sin\theta_{\rho}\,
\sin\theta_{\lambda}
\nonumber \\
&=& -\left( \,\frac{4\,\pi}{\sqrt{3}}\right)
\left[Y_{1}(\Omega_{\rho}) \otimes Y_{1}(\Omega_{\lambda})
\right]_{L=0}
\end{eqnarray}
The infinitesimal volume element $dV = R^{5} dR d\Omega ^{(5)}$
and the infinitesimal hyper-spherical solid angle $d\Omega ^{(5)}$
are given by
\begin{eqnarray}
dV &=& R^{5} dR d\Omega ^{(5)},\\
\quad d\Omega ^{(5)} &=& \cos ^{2}\chi \, \sin ^{2} \chi \,d\chi
\,\sin \theta \,d\theta
\,d\Omega ^{(3)}, \\
\quad d\Omega ^{(3)} &=&
\,d\phi \,d\Omega ^{(2)}, \\
\quad d\Omega ^{(5)} &=& \cos ^{2}\chi \, \sin ^{2} \chi \,d\chi
\,d\Omega_{\rho} ^{(2)} \,d\Omega_{\lambda} ^{(2)}, \
\end{eqnarray}
where $d\Omega ^{(2)}$ is the usual three-dimensional space
differential solid angle. Then one has
\begin{equation}
\int d\Omega ^{(2)} = 4 \pi, \quad \int d\Omega ^{(3)} = 8 \pi
^{2}, \quad \int d\Omega ^{(5)} = \pi^{3}.
\end{equation}
The ground state matrix element $\langle V_{\rm Y}\rangle_{00}(R)$
is often called the hyper-radial potential $V_{\rm Y}(R)$; the
integrals in the hyper-radial Y-string potential $V_{\rm Y}(R)$
can be evaluated numerically as
\begin{eqnarray}
V_{\rm Y}(R) = \langle V_{\rm Y}(R)\rangle_{00} &=& \int~ {\cal
Y}_{0,0,0}^{*}(\Omega_{\lambda}) V_{\rm Y}({\bm \rho }, {\bm
\lambda })\, {\cal Y}_{0,0,0}(\Omega_{\rho}) \, d\Omega ^{(5)}
\nonumber \\
&=& \frac{1}{\pi ^{3}} \int V_{\rm Y}({\bm \rho }, {\bm \lambda
})\, d\Omega ^{(5)},
\nonumber \\
&=& \frac{8}{\pi } \int_{0}^{\pi/2} \cos ^{2}\chi \, \sin^{2}\chi
\,d\chi \, \int_{0}^{\pi} V_{\rm Y}(R,\chi, \theta )\,\sin \theta
\,d\theta
\nonumber \\
&=&  \sigma \sqrt{\frac{3}{2}} R \frac{8}{\pi }\,\int_{0}^{\pi}
\,\sin\theta \,d\theta \int_{0}^{\frac{\pi}{2}} {\sqrt{1 +
|\sin\theta| \,\sin2\,\chi}} \,{\left(\cos\chi\,\sin\chi
\right)}^2\, d\chi \nonumber \\
&\simeq& 1.2876 \, \sigma \sqrt{\frac{3}{2}} R .\
\label{vhpc1}
\end{eqnarray}
where $|\sin \theta | = \sqrt{1 - \cos^{2}\theta}.$ One can easily
write down the power expansion of the three-body potential $V_{\rm
Y}$; thus one obtains an expansion of the potential in (integer)
powers of ordinary spherical harmonics and of $\sin2\chi$, which
can be combined into (new) hyper-spherical harmonics ${\cal
Y}_{K,L,m_{i}}$ that can be used in Eqs. (\ref{vhpc}),
(\ref{vhpc1}) by using the (hyper-spherical) Clebsch-Gordan
coefficients, or simply by brute-force numerical integration. Note
that these matrix elements are ``almost always" diagonal, i.e.,
that this potential does not connect hyper-spherical harmonics
with different K, but only of the same K and L, though possibly
with different $l_{\rho}$ and $l_{\lambda}$ values, due to
properties of the three-dimensional space spherical harmonics.

\subsection{K=0,1,2 hyper-spherical harmonics}
\label{s:K012}

Manifestly, in order to be able to evaluate Eq. (\ref{vhpc}) one
must know the explicit form of the hyper-spherical harmonics.
Simonov, Ref. \cite{Simonov:1965ei}, has explicitly written down
the $K=0,1,2$ states' wave functions (hyper-spherical harmonics)
and examined their permutation symmetry properties, although the
higher-K symmetrized hyper-spherical harmonics have remained a
widely unexplored topic \cite{Krivec}.

The symmetries of the string potential/hamiltonian are: parity,
permutation/spatial exchange of quarks, rotation, therefore we see
that only wave functions with the same $P=(-1)^{l_1+l_2}$, $L$,
and symmetry $M,S,A$ can mix with each other. Let $P_{ij}$ be the
$ij$-th particle permutation/spatial exchange operator. The
permutation symmetry can be examined using the following
properties of the $(\rho,\lambda)$ set of vectors with mixed
symmetry \cite{richard92}
\begin{subequations}
\begin{eqnarray}
P_{12}\vec{\rho} & \rightarrow -\vec{\rho} \\
P_{12}\vec{\lambda} & \rightarrow \vec{\lambda} \\
P_{13}\vec{\rho} & \rightarrow \frac{1}{2}\vec{\rho} -
 \frac{\sqrt{3}}{2}\vec{\lambda} \\
P_{13}\vec{\lambda} & \rightarrow -\frac{\sqrt{3}}{2}\vec{\rho} -
 \frac{1}{2}\vec{\lambda}
\end{eqnarray}
\end{subequations}
that furnish the basis for the two-dimension irrep of $S_3$. Using
the above relations, the second-order scalar, vector, and tensor
can be constructed and used to obtain the $K = 0,1,2$
hyper-spherical harmonics, see Ref. \cite{Simonov:1965ei}. Thus,
it turns out that the $S_3$ permutation group symmetrized
hyper-spherical harmonics correspond precisely to different
SU(6)$_{FS}$ symmetry multiplets (Young diagrams/tableaux) of the
three-quark system. The symmetrized hyper-spherical harmonics have
been systematically developed only up to K=2, see comments in Ref.
\cite{Krivec}. We show them in Table \ref{tab:hypsphharm}.
\begin{table}[tbh]
\begin{center}
\caption{The hyper-spherical harmonics, for K = 0,1,2 values.}
\begin{tabular}{ccccccccc}
\hline \hline K & $n_K$ & $[SU_{FS}(6),L^P]$ & $(l_{\rho} \times
l_{\lambda})_{L}$ & hyper-spherical harmonic \\
\hline
0 & 1 & $[56,0^+]$ & $(0 \times 0)_{0}$ & ${\cal Y}_1 =
\frac{4}{\sqrt{\pi}} [Y_0 \otimes Y_0]_0$ \\
\hline
1 & 3 & $[70,1^-]$ & $(1 \times 0)_{1}$ & ${\cal Y}_{2a} =
4 \sqrt{\frac{2}{\pi}}\cos\chi
[Y_1 \otimes Y_0]_1$  \\
 &  & $[70,1^-]$ & $(0 \times 1)_{1}$ & ${\cal Y}_{2b} = 4
\sqrt{\frac{2}{\pi}}\sin\chi [Y_0 \otimes Y_1]_1$ \\
\hline 2 & 1 & $[70,0^+]$ & $(0 \times 0)_{0}$ & ${\cal Y}_{3a} =
\frac{8}{\sqrt{\pi}}(\cos\chi^2 - \sin\chi^2) [Y_0 \otimes Y_0]_0$ \\
 &  & $[70,0^+]$ & $ (1 \times 1)_{0}$ & $ {\cal Y}_{3b} =
\frac{16}{\sqrt{3\pi}} \cos\chi \sin\chi [Y_1 \otimes Y_1]_0$  \\
\hline
2 & 3 & $[20,1^+]$ & $(1 \times 1)_{1}$ & ${\cal Y}_{4} =
\frac{16}{\sqrt{3\pi}} \cos\chi \sin\chi [Y_1 \otimes Y_1]_1$ \\
\hline 2 & 5 & $[56,2^+]$ & $\begin{array}{c}~~(2 \times 0)_{2} \\
+ (0 \times 2)_{2}
\end{array}$ & ${\cal Y}_{5} =
\frac{16}{\sqrt{5\pi}}\frac{1}{\sqrt{2}}\left[ \cos^2\chi [Y_2
\otimes Y_0]_2 + \sin^2\chi [Y_0 \otimes Y_2]_2
\right]$ \\
\hline 2 & 5 & $[70,2^+]$ & $\begin{array}{c}~~(2 \times 0)_{2} \\
- (0 \times 2)_{2}
\end{array}$
& ${\cal Y}_{6a} = \frac{16}{\sqrt{5\pi}}\frac{1}{\sqrt{2}}\left[
\cos^2\chi [Y_2 \otimes Y_0]_2 - \sin^2\chi [Y_0 \otimes Y_2]_2 \right]$ \\
 &  & $[70,2^+]$ & $(1 \times 1)_{2}$ & ${\cal Y}_{6b} =
\frac{16}{\sqrt{3\pi}} \cos\chi \sin\chi [Y_1
\otimes Y_1]_2$ \\
\hline
\end{tabular}
\label{tab:hypsphharm}
\end{center}
\end{table}

\section{The hyper-angular matrix elements}
\label{s:Y-HS ME}

\subsection{First-order expansion of the Y-string potential}
\label{s:HSang}

The following is, of course, just a schematic representation of
the total matrix elements, as the complete K=2 band wave functions
may have two components with different partial orbital momentum
components $\left(l_{\rho}, l_{\lambda}\right)$ and identical
total orbital angular momentum $L$, such as the K=2 band
multiplets $[70,0^+]$ and $[70,2^+]$ (see above):
\begin{eqnarray}
\langle \, K (L_{f}) \, |\,\sin2\chi \,|\sin \theta | \, |\,
K(L_{i}) \, \rangle_{\rm hyp-ang} &=& \langle \,
\psi_{K(L_{f})}(\chi) \, | \,\sin2\chi \,|\, \psi_{K(L_{i})}(\chi)
\, \rangle_{\rm hyp} \langle \, Y_{L_{f}} \,|
\,|\sin \theta \, | \,|\, Y_{L_{i}} \, \rangle_{\rm ang} \nonumber \\
&=& \sum_{l_{1},l_{2}} \langle \, \psi_{K(L_{f})}(\chi) \, |
\,\sin2\chi \, | \, \psi_{K(L_{i})}(\chi) \, \rangle_{\rm hyp} \nonumber \\
&\times& \langle \left(l_{1f} \otimes
l_{2f}\right)_{L_{f}}^{M_{f}}| \,|\sin \theta | \, |\left(l_{1i}
\otimes l_{2i}\right)_{L_{i}}^{M_{i}} \rangle_{\rm ang} .
\end{eqnarray}
We may drop the indices for the initial and the final state total
orbital angular momenta $L_{f}=L_{i}=L; M_{f}=M_{i}=M$, as 
the angular matrix elements are diagonal in those quantum numbers.
We use the spatial wave functions available in Refs.
\cite{FH69,IK79,Simonov:1965ei} to calculate the hyper-angular
matrix elements $\langle \, \psi_{\rm K} \, | \, \frac{2\, \rho
\,\lambda}{R^{2}} \, |\, \psi_{\rm K} \, \rangle_{\rm hyp-ang} =
\langle \, \psi_{\rm K} \, | \,  \sin2\chi \, |\, \psi_{\rm K} \,
\rangle_{\rm hyp-ang}$ and take the angular matrix elements
$\langle \, Y_L \, |\,|\sin \theta |\, |\, Y_L \, \rangle_{\rm
ang}$ from Table \ref{tab:drcc1a} to calculate the total
hyper-angular matrix element $\langle \, \psi_K \, |\,
|\,\sin2\chi \,|\sin \theta |\, |\, \psi_K \, \rangle_{\rm
hyp-ang}$. We list the radial-, angular- and total matrix
elements' values, as well as the results of some intermediate
steps, in Table \ref{tab:drcc2mB}.
\begin{table}[tbh]
\begin{center}
\caption{The values of the three-body potential hyper-angular
matrix elements $\langle \frac{2\, \rho \lambda}{R^2}\rangle_{\rm
hyp} = \langle \sin2\chi \, \rangle_{\rm hyp}$ and $\langle
\sin2\chi \,|\sin \theta| \rangle_{\rm hyp-ang}$, for the K =
0,1,2 states (in all the partial S-, P- and D-waves). In the
$[56,2^+]$ and the $[70,2^+]$ entries ``diag." denotes diagonal
terms of the type $(2 \times 0)_{2} \times (2 \times 0)_{2}$ and
$(0 \times 2)_{2} \times (0 \times 2)_{2}$, whereas ``off-diag."
denotes off-diagonal terms $(2 \times 0)_{2} \times (0 \times
2)_{2}$ and $(0 \times 2)_{2} \times (2 \times 0)_{2}$. These two
kinds of contribution must be separated because they have
different angular matrix elements $\frac{4}{\pi}\langle |\sin
\theta| \rangle$.}
\begin{tabular}{ccccccccc}
\hline \hline K & $[SU_{FS}(6),L^P]$ & $(l_{\rho} \times
l_{\lambda})_{L}$ & N & $\langle \frac{2\, \rho
\lambda}{R^2}\rangle$ & $\frac{3\,{\pi}}{8}\langle \sin2\chi \,
\rangle$ & $\frac{4}{\pi}\langle |\sin \theta| \rangle$ & $\langle
\sin2\chi \,|\sin \theta| \rangle$ & $=$ \\
\hline
0 & $[56,0^+]$ & $(0 \times 0)_{0}$ & 3 & $\frac{8}{3\pi}$
& $1$ & $1$ & $\frac{2}{3}$ & 0.6667 \\
1 & $[70,1^-]$ & $(1 \times 0)_{1}$ & 4 & $\frac{8}{3\pi}$ & $1$ &
$1$ & $\frac{2}{3}$ & 0.6667 \\
0 & $[56,0^+]$ & $ (0 \times 0)_{0}$ & 5 & $\frac{8}{3\pi}$ & $1$
& $1$ & $\frac{2}{3}$ & 0.6667 \\
2 & $[20,1^+]$ & $(1 \times 1)_{1}$ & 5 & $\frac{128}{45\,{\pi}}$ &
$\frac{16}{15}$ & $\frac{9}{8}$ & $\frac{4}{5}$ & 0.8000 \\
\hline
2 & $[70,0^+]$ & $ (0 \times 0)_{0}$ & 5 & $\frac{32}{15 \,{\pi}}$ &
$\frac{4}{5}$ & $1$ & $\frac{8}{15}$ & 0.5333 \\
2 & $[70,0^+]$ & $ (1 \times 1)_{0}$ & 5 & $\frac{128}{45\,{\pi}}$
& $\frac{16}{15}$ & $\frac{3}{4}$ & $\frac{8}{15}$ & 0.5333 \\
\hline 2 & $[56,2^+]$ & diag. & 5 & $\frac{64}{25\,{\pi}}$ &
$\frac{24}{25}$ & $1$ & $\frac{16}{25}$ & 0.6400 \\
2 & $[56,2^+]$ & off-diag. & 5 & $\frac{32}{75{\pi}}$ &
$\frac{4}{25}$ & $-\frac{1}{8}$ & $-\frac{4}{75}$ & -0.0533 \\
2 & $[56,2^+]$ & $\begin{array}{c}~~(2 \times 0)_{2} \\  + (0
\times 2)_{2}
\end{array}$ & 5 & & & & $\frac{44}{75}$ & 0.5867 \\
\hline 2 & $[70,2^+]$ & diag. & 5 &
$\frac{64}{25\,{\pi}}$ & $\frac{24}{25}$ & $1$ & $\frac{16}{25}$ & 0.6400 \\
2 & $[70,2^+]$ & off-diag. & 5 & $-\frac{32}{75{\pi}}$ &
$-\frac{4}{25}$ & $-\frac{1}{8}$ & $\frac{4}{75}$ & 0.0533 \\
2 & $[70,2^+]$ & $\begin{array}{c}~~(2 \times 0)_{2} \\  - (0
\times 2)_{2}
\end{array}$ & 5 & & & & $\frac{52}{75}$ & 0.6933 \\
2 & $[70,2^+]$ & $(1 \times 1)_{2}$ & 5 & $\frac{128}{45 \,{\pi}}$
& $\frac{16}{15}$ & $\frac{39}{40}$ & $\frac{52}{75}$ & 0.6933 \\
\hline
\end{tabular}
\label{tab:drcc2mB}
\end{center}
\end{table}

\subsection{The Y-string potential}
\label{s:NPYstring}

The total matrix elements may have two components with different
partial orbital momenta $\left(l_{\rho}, l_{\lambda}\right)$ and
identical total orbital angular momentum $L$, e.g. the K=2 band
multiplets $[70,0^+]$ and $[70,2^+]$: where
\begin{eqnarray}
\langle \, \psi_{K(L_{f})}(\chi) \,| \,\sqrt{1 + \sin2\chi
|\sin\theta|} \,|\, \psi_{K(L_{i})}(\chi) \, \rangle_{\rm hyp} &=&
\int_{0}^{\frac{\pi}{2}} \left(\frac{1}{2}\sin2\chi\right)^2 \,
|\psi_{K,L}(\chi)|^2\, \sqrt{1 + \sin2\chi |\sin\theta|}\,d\chi .
\label{e:int1}
\end{eqnarray}
and we are left with a double integral over the product of
even-$L$ order Legendre polynomials $P_{L}(x)$, the potential
$\sqrt{1 + \sin2\chi |\sin\theta|}$ and the hyper-angular wave
function squared $|\psi_{K,L}(\chi)|^2$ :
\begin{eqnarray}
\frac{1}{2}\int_{-1}^{1}d\cos\theta P_L(\cos \theta) \,\langle \,
\psi_{K(L_{f})} \,| \,\sqrt{1 + \sin 2 \chi |\sin \theta|} \,|\,
\psi_{K(L_{i})} \, \rangle &=& \frac{1}{2}\,
\int_{-1}^{1}d\cos\theta P_L(\cos \theta) \,
\int_{0}^{\frac{\pi}{2}}d\chi \,
\left(\frac{1}{2}\sin2\chi\right)^2 \,
\nonumber \\
&\times& |\psi_{K,L}(\chi)|^2\, \sqrt{1 + \sin2\chi |\sin\theta|}
. \label{e:int2}
\end{eqnarray}
We use the K=0,1,2 hyper-angular wave functions available in
Appendix \ref{s:K012} to calculate these hyper-angular matrix
elements and take the angular momentum coefficients from Table
\ref{tab:drcc1a}. The integral can be rewritten in terms of new
variables $z = \cos2\chi$ and $x=\cos\theta$ as follows
\begin{eqnarray}
\frac{1}{2}\int_{-1}^{1}d\cos\theta P_L(\cos \theta) \,\langle \,
\psi_{K(L_{f})}(\chi) \,| \,\sqrt{1 + \sin 2 \chi |\sin \theta|}
\,|\, \psi_{K(L_{i})}(\chi) \, \rangle_{\rm hyp} &=& \frac{1}{2}\,
\int_{-1}^{1}dx P_L(x) \,\left(\frac{1}{2}\right)^3
\,\int_{-1}^{1}\sqrt{1 - z^2}dz
\nonumber \\
&\times& |\psi_{K,L}(z)|^2\, \sqrt{1 + \sqrt{1 - z^2} \sqrt{1 -
x^2}} , \label{e:int3}
\end{eqnarray}
and evaluated numerically; we list the results in Table
\ref{tab:drcc2m}.
\begin{table}[tbh]
\begin{center}
\caption{The values of the three-body potential hyper-angular
matrix elements $\langle \sqrt{1 + \sin 2 \chi |\sin \theta|}
\rangle_{\rm hyp-ang.}$ for the K = 0,1,2 states (in all the
partial S-, P- and D-waves).}
\begin{tabular}{ccccccccc}
\hline \hline K & $[SU_{FS}(6),L^P]$
& $\langle \sqrt{1 + \sin 2 \chi |\sin \theta|} \rangle_{\rm hyp-ang.}$ \\
\hline 0 & $[56,0^+]$ & 1.2876 \\
1 & $[70,1^-]$ & 1.2876 \\
0 & $[56,0^+]$ & 1.2876 \\
2 & $[20,1^+]$ & 1.3402 \\
2 & $[70,0^+]$ & 1.2350\\
2 & $[56,2^+]$ & 1.2560 \\
2 & $[70,2^+]$ & 1.2981 \\
\hline
\end{tabular}
\label{tab:drcc2m}
\end{center}
\end{table}

\subsection{The complete string potential}
\label{s:NPCompletestring}

\begin{figure}[tbp]
\centerline{\includegraphics[width=4.5in,,keepaspectratio]{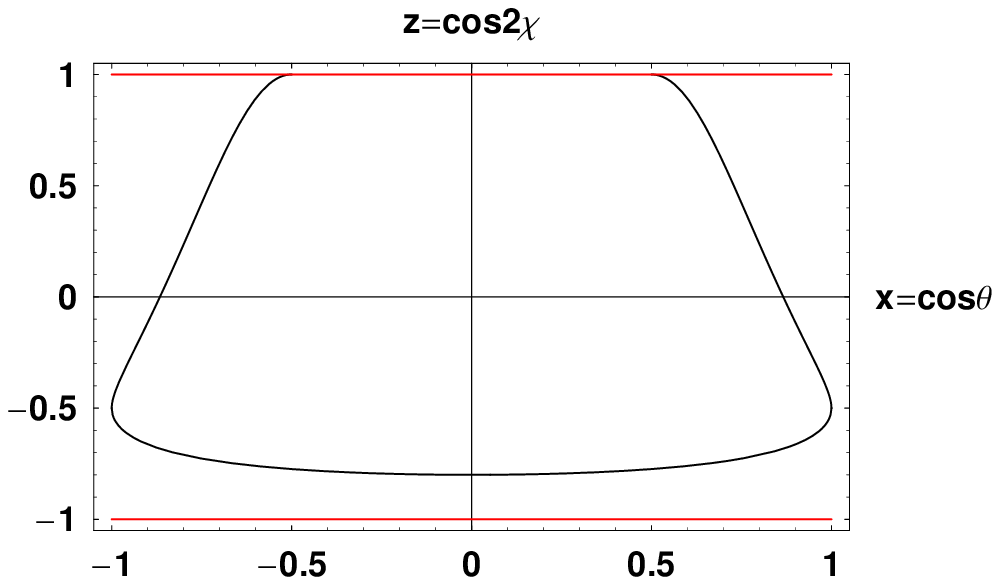}}
\caption{The boundary in the $z = \cos2\chi$ vs. $x=\cos\theta$
plane, between the regions in which the two- and the three-string
potentials are appropriate, see Eqs. (\ref{e:boundary2}).}
\label{f:bound2}
\end{figure}

The boundary Eqs. (\ref{e:boundary}) can be recast in the new
variables $z = \cos2\chi$ and $x=\cos\theta$ of Fabre de la
Ripelle and Lassaut, Ref. \cite{fabr97}, that are particularly
useful in the integration over the solid hyper-angle, see Fig.
\ref{f:bound2}:
\begin{eqnarray}
z_1(x) &=& \frac{5\,x^2 - 8\,x^4 + {\sqrt{3}}\,x\,{\sqrt{1 -
x^2}}}{1 - 3\,x^2 + 8\,x^4 + {\sqrt{3}}\,x\,{\sqrt{1 - x^2}}},
\nonumber \\
z_2(x) &=& \frac{-5\,x^2 + 8\,x^4 + {\sqrt{3}}\,x\,{\sqrt{1 -
x^2}}}{-1 + 3\,x^2 - 8\,x^4 + {\sqrt{3}}\,x\,{\sqrt{1 - x^2}}},
\nonumber \\
z_3(x) &=& \frac{2 + x^2 + {\sqrt{4 - 5\,x^2 + x^4}}}{-7 + x^2 +
{\sqrt{4 - 5\,x^2 + x^4}}}. \label{e:boundary2} \
\end{eqnarray}
Due to the symmetry of the integrand we may take twice the
integral over $x$ from 0 to 1; moreover the integral can be
rewritten in terms of the new variables $z = \cos2\chi$ and
$x=\cos\theta$ as follows
\begin{eqnarray}
\frac{1}{2}\int_{-1}^{1}d\cos\theta P_L(\cos \theta) \, \langle \,
\psi_{K(L_{f})}(\chi) \,| \,V(\chi,\theta) \,|\,
\psi_{K(L_{i})}(\chi) \, \rangle_{\rm hyp} &=&~\left(
\int_{0}^{\frac12}dx \int_{z_3(x)}^{1}\, dz + \int_{\frac12}^{1}dx
\int_{z_3(x)}^{z_1(x)}\, dz \right)
\nonumber \\
\left(\frac{1}{2}\right)^3 \,P_L(x) \,\sqrt{1 - z^2} \,
&\times&
|\psi_{K(L)}(z)|^2 \,
\sqrt{\frac32} \sqrt{1 + \sqrt{1 - z^2} \sqrt{1 - x^2}}
\nonumber \\
&\times& \left(\sqrt{1 + \frac{z}{2} + \frac{x}{2}\sqrt{3(1 -
z^2)}}
+ \sqrt{1 + \frac{z}{2} - \frac{x}{2}\sqrt{3(1 - z^2)}}\right) \nonumber \\
&+& \left(\frac{1}{2}\right)^3 \, \int_{\frac12}^{1}P_L(x) \,dx
\int_{z_1(x)}^{1}|\psi_{K(L)}(z)|^2\,\sqrt{1 - z^2}dz
\nonumber \\
&\times& \left(\sqrt{1 - z} + \sqrt{1 + \frac{z}{2} -
\frac{x}{2}\sqrt{3(1 - z^2)}}\right),\label{e:int4}
\end{eqnarray}
where $z_{1,3}(x)$ are given by Eqs. (\ref{e:boundary2}) defining
the boundary in the right-hand-side half of the $x$ vs. $z$ plane.
\begin{table}[tbh]
\begin{center}
\caption{The values of the three-body potential hyper-angular
matrix elements $\langle V_{\rm string} \rangle_{\rm hyp-ang.}$
for the K = 0,1,2 states (in all the partial S-, P- and D-waves).}
\begin{tabular}{ccccccccc}
\hline \hline K & $[SU_{FS}(6),L^P]$
& $\langle V_{\rm string} \rangle_{\rm hyp-ang.}$ \\
\hline 0 & $[56,0^+]$ & 1.289126 \\
1 & $[70,1^-]$ & 1.289126 \\
0 & $[56,0^+]$ & 1.289126 \\
2 & $[20,1^+]$ & 1.340371 \\
2 & $[70,0^+]$ & 1.240090 \\
2 & $[56,2^+]$ & 1.258379 \\
2 & $[70,2^+]$ & 1.298492 \\
\hline
\end{tabular}
\label{tab:drcc2m2}
\end{center}
\end{table}

\section{The hyper-radial Schr\" odinger equation}
\label{s:Hrad}

\subsection{Solving the hyper-radial equation}
\label{s:Hrad2}

The eigenvalues of the three quark states can be obtained by
solving the eigenvalue problem of the following equation
\begin{eqnarray}
\left[\frac{d^2}{dR^2} + \frac{5}{R}\frac{d}{dR} -
\frac{K(K+4)}{R^2} + 2 \mu E - 2\mu\sqrt{\frac{3}{2}}\sigma R
\right]\psi_{[K]} = 0
\end{eqnarray}
By changing scale of the hyper-radial coordinate one can write
\begin{eqnarray}
\left[\frac{d^2}{dx^2} + \frac{5}{x}\frac{d}{dx} -
\frac{K(K+4)}{x^2} + a - x \right]\psi_{[K]} = 0
\end{eqnarray}
with
\begin{eqnarray}
R &  = & \alpha x \\
\alpha & = & \left(2\mu \sqrt{\frac{3}{2}}\sigma \right)^{-1/3} \\
a & = & 2 \mu E \alpha^2
\end{eqnarray}
Regular solution can be written $\psi_{[K]} = x^K u_{[K]}(x)$ with
$u_{[K]}(0) = const$. The equation for $u_{[K]}$ can be written as
\begin{eqnarray}
\left[\frac{d^2}{dx^2} + \frac{2K + 5}{x}\frac{d}{dx} + a - x
\right]u_{[K]} = 0
\end{eqnarray}
The above equation is solved with the boundary condition
\begin{eqnarray}
u(0) & = & \mbox{finite} \\
u(\infty) & \rightarrow 0
\end{eqnarray}
The obtained eigenvalues are tabulated in Table \ref{t:energies1}.
\begin{table}
\begin{tabular}{c|ccc}\hline
 $K$ & $a_0$ & $a_1$ & $a_2$ \\ \hline
 0 & 3.82 & 5.26  & 6.54 \\
 1 & 4.66 & 5.99  & 7.19 \\
 2 & 5.43 & 6.67  & 7.81
\end{tabular}
\caption{Energy eigenvalues of the three-quark states in the
linear hyper-radial potential for the ground- ($a_0$), the first-
($a_1$) and the second ($a_2$) radially excited states, in natural
units (see text).} \label{t:energies1}
\end{table}

\subsection{Evaluation of the potential's hyper-radial matrix element}
\label{s:Hrad1}

We do not need the hyper-radial wave functions in order to
complete the calculation of the radial matrix elements; rather, we
use the virial theorem
$$V(k)=\frac{2}{k+2}E(k)$$
to determine the expectation value $\langle V(k=1) \rangle
=\frac{2}{3}E(k=1)$, where $k$ is the power of $r$ in the
potential $V(k) \sim r^{k}$, in terms of the energy eigenvalue
$E(k=1)$ of the unperturbed hamiltonian.

\end{document}